\documentclass[preprint]{aastex62}

\usepackage{enumitem,amssymb}
\newlist{todolist}{itemize}{2}
\setlist[todolist]{label=$\square$}
\usepackage{pifont}

\usepackage{soul}
\usepackage{colortbl}

\graphicspath{{./}{figures/}}

\received{October 15th, 2019}
\revised{March 2nd, 2020}
\accepted{March 24th, 2020}
\submitjournal{ApJ}

\shorttitle{Meteoroid Bombardment of Lunar Poles}
\shortauthors{Pokorn\'{y} et al.}

\begin{document}

\title{Meteoroid Bombardment of Lunar Poles}

\correspondingauthor{Petr Pokorn\'{y}}
\email{petr.pokorny@nasa.gov}

\author{Petr Pokorn\'{y}}
\affiliation{Department of Physics, The Catholic University of America, Washington, DC 20064, USA}
\affiliation{Heliophysics Science Division, NASA Goddard Space Flight Center, Greenbelt, MD 20771}
\affiliation{Astrophysics Science Division, NASA Goddard Space Flight Center, Greenbelt, MD 20771}
\author{Menelaos Sarantos}
\affiliation{Heliophysics Science Division, NASA Goddard Space Flight Center, Greenbelt, MD 20771}
\author{Diego Janches}
\affiliation{Heliophysics Science Division, NASA Goddard Space Flight Center, Greenbelt, MD 20771}
\author{Erwan Mazarico}
\affiliation{Solar System Exploration Division, NASA Goddard Space Flight Center, Greenbelt, MD 20771}

\begin{abstract}
 While the floors of deep lunar craters are largely shielded from solar radiation and thus provide an ideal thermal environment for water ice accumulation,  meteoroids on highly inclined orbits can easily access permanently shadowed regions and alter the surface properties via hyper-velocity impacts. Here we consider the detailed topography of lunar poles and a dynamical model of meteoroids to quantify the meteoroid mass fluxes, energy deposition, and impact ejecta mass production rates. 
  Our analysis of regions within $5^\circ$ from the two lunar poles shows that the variations of the meteoroid mass flux, energy flux and ejecta production rate are within 50\% of their median values. We find that lunar poles are easily accessible by meteoroid impacts including permanently shadowed regions. We find a positive correlation between the surface slope and the meteoroid ejecta production rate, a finding that suggests a higher impact gardening rate on steep crater walls can facilitate mass wasting.

\end{abstract}

\section{Introduction}
The polar regions of the Moon are interesting both from a scientific perspective, because they can provide clues to physical processes that apply to the entire Solar System, as well as for human exploration, because they may contain possible space exploration resources in the form of volatiles. High resolution surface temperature maps of both lunar poles by the Lunar Reconnaissance Orbiter (\textit{LRO}) Diviner instrument \citep{Paige_etal_2010}, combined with information on the topography derived from Lunar Orbiter Laser Altimeter \textit{(LOLA)} observations \citep{Smith_etal_2010}, provide a unique way of characterizing permanently shadowed regions that consistently retain low temperatures and thus offer stability for water-ice deposits on the Moon over geological timescales. \citet{HAYNE_ETAL_2015}, \citet{FISHER_ETAL_2017}, \citet{Li_Milliken_2017} \added{ and \citet{Li_etal_2018}} suggested that \added{many of} the south pole permanently shadowed regions (PSRs) exhibit signatures of exposed water-ice deposits\added{, while other PSRs lack the water-ice signatures completely}. \added{\citet{Rubanenko_etal_2019}, using small crater morphology and distribution on Mercury and Moon, suggested that water-ice is present in significantly more lunar south pole craters than previously thought from observations \citep{Colaprete_etal_2010,HAYNE_ETAL_2015}}. However, the distribution of water-ice layers was not correlated with the maximum surface temperature, where some regions such as the Shackleton or Shoemaker craters showed  either anomalously low or no amount of water-ice. Due to shielding from the solar wind, the most significant exogenous sources of depletion of surface ice are expected to be \deleted{either}\added{local insterstellar H Ly-$\alpha$ radiation \citep{Morgan_Shemansky_1991}}, mass wasting on highly inclined slopes \citep{FISHER_ETAL_2017}, meter-size or larger impacts \citep{Suggs_etal_2014}, and  impacts of smaller meteoroids that are constantly bombarding the lunar poles with a broad range of impact velocities \citep{Pokorny_etal_2019}.
Recent work by~\citet{Deutsch_etal_2019} further supports a very patchy spatial distribution of surface ice, which suggests high overturn or destruction rates. 

The objective of this work is to evaluate how the access of smaller meteoroids changes from crater to crater at high lunar latitudes. This calculation provides critical data to test whether albedo differences and/or the patchiness seen in surface ice signatures might be attributed to differences in meteoroid impacts. Whereas the effect of impactors in the centimeter to meter scale on the gardening rate has been quantified \citep{Hurley_etal_2012,Gault_1973,Arnold_1975}, the effects of sporadic meteoroids, which are smaller but far more frequent, on the topographically complex polar regions of the Moon have not been quantified.
 The first high resolution dynamical model of meteoroids impacting the lunar surface,  presented by \citet{Pokorny_etal_2019}, provided an overview of the different physical effects that meteoroid impacts produce on the lunar surface, yet a major limitation of that investigation was its assumption of a spherical Moon geometry. Here, we expand this work by combining the  meteoroid direction, velocity, mass and flux distribution from the \citet{Pokorny_etal_2019} model with the detailed topography of lunar poles to evaluate the meteoroid flux, the total energy delivered by meteoroids, and the ejecta mass created through meteoroid impacts of micron to mm size particles. 

\section{Lunar datasets and Meteoroid model}
In this manuscript we focus on both lunar polar caps, specifically the region limited to selenographic latitudes between $85^\circ-90^\circ$ (approximately 150 km from the north/south pole). For this study we use the LOLA GDR (Gridded Data Records) topography map and the sporadic meteoroid background model to quantify the effects of meteoroid impacts onto the lunar poles. Ancillary data used for this work were the Diviner depth of ice, maximum temperature, and average temperature and LOLA albedo. The links and names of all data sets used here are described in Appendix \ref{Sec:AppendixA}.

We transformed the original LOLA GDR data set into a $250$ m $\times$ $250$ m Cartesian grid in $X,Y$, with the $Z$ component  obtained using bi-linear interpolation applied to the LOLA altimeter data set. The Diviner maximum temperature, average temperature, depth of ice, and LOLA albedo maps were obtained in the form of triangles (i.e., set of values for each triangle on the surface mesh). For each grid point in our 250 m Cartesian grid we found the relevant Diviner surface triangle and its corresponding values for our calculation (see the project GitHub page for the codes used in this manuscript). For missing values we filled the blanks with a bi-linear interpolation using four points closest to the missing grid point. 
Surface slopes were calculated from our gridded data set by slicing the grid squares, denoted by four vertices, into two triangles and calculating the slope as the average of the two, also keeping the triangle slope difference. For grid squares with differences higher than $0.1^\circ$, we added two more triangles (sliced in the reverse direction) and used the average of all four triangle slopes as the final value.
We tested the average slope calculated using our method against the LOLA data set and $99.99\%$ of our slopes were $<0.1\%$ different from the LOLA slope data set.

The lunar meteoroid environment model used in this work was described in detail in \citet{Pokorny_etal_2019}. In this model the inner solar system is populated by meteoroids originating in four distinctive populations: main-belt asteroids (MBA), Jupiter family comets (JFC), Halley-type comets  (HTC) and Oort Cloud comets (OCC). We considered particle diameters between $D=10~\mu$m and $D=2,000~\mu$m, with an assumed bulk density equal to 3 g cm$^{-3}$ for the main-belt meteoroids, and 2 g cm$^{-3}$ for cometary meteoroids. The model was calibrated with the mass influx at Earth from \citet{CarrilloSanchez_etal_2016} yielding a yearly average of 1.4 metric tons a day of material delivered onto the lunar surface from asteroidal and short-and-long cometary sources. To simulate meteoroid effects on the scale of millions of years, we calculated  moments of the meteoroid input function from this model  by accumulating model output over one Earth orbit around the Sun from 00:00:00 UTC, July 1st, 2013 to 12:00:00 UTC June 30th, 2014. These calculations  combined 731 temporal slices (each 12 hours apart from the next) into one average file.

 Based on \citet{Pokorny_etal_2019} we expect negligible changes in the annual average for different years due to the fact that  Earth's heliocentric distance variations, which are the main contributor to the variations of the meteoroid flux at the Moon,  are negligible from year-to-year.   Because in this calculation we have marginalized the meteoroid differences experienced by a crater at different local times and months of the year, the only parameter influencing the year-to-year comparisons are the amplitudes of monthly variations due to the orbital motion of the Moon around Earth, i.e., the lunar phase at the start of a  year. \citet{Pokorny_etal_2019} showed that the highest meteoroid flux occurs  close to the full moon, while the minimum occurs around the new moon. 
 The most favorable configuration (all monthly maxima are counted within the year while the last monthly minimum  is not counted) versus the least favorable configuration leads to a difference of approximately 3.5\% in the mass flux averaged over one year \citep[using Eq. 12 in ][]{Pokorny_etal_2019}.

\section{Methods}
 In order to map the meteoroid impacts onto the topography of the lunar poles, we binned the meteoroid radiant/sky map in  ecliptic longitude (2 degree bins), latitude (2 degree bins), and the meteoroid impact velocity at the Moon (2 km s$^{-1}$ bins). Each of these 3D bins provides  the yearly averaged mass flux, 
 direction, and velocity allowing us to calculate the incident mass flux on each surface segment on the Moon. For simplicity we assume that the Moon has no axial tilt to avoid incorporating seasonal variations, which would increase the already computationally demanding processing part. Due to the broad range of meteoroid ecliptic latitudes, this simplification  has a negligible effect on the results shown here. The small fraction of the total meteoroid flux that is very close to the ecliptic, and their shallow impact angles with respect to the polar caps, diminish the overall contribution of these low-inclination impactors.

For each surface element (triangle) we first determine whether the surface element is  obstructed by any  topographic feature with respect to the particular radiant point using  ray-tracing methods. First, we rotate all surface elements with respect to the center of the Moon  such that the incident meteoroid ray is originating from $(0,0,Z)$ direction, i.e., the ray is only traveling in the $Z$-direction. Then we select all the triangles that are able to shadow the inspected surface element using bounding boxes (i.e., rectangles defined by the minimum and maximum $x$ and $y$ values of each triangle) 
and the $Z$ component of all surface elements. Then we check whether the centroid of the inspected surface element is shadowed by any of the triangles from the selected set. This procedure results in shadow maps of both polar regions for all directions available from our meteoroid model. We tested our shadowing procedure against the algorithm presented in \citet{Mazarico_etal_2018} and obtained essentially identical results. 

Using the shadow maps we then determine the mass flux $\mathcal{M}$, energy flux $\mathcal{E}$, and ejecta mass production rate $\mathcal{P}^+$ for each surface element using the following expressions:



\begin{eqnarray}
\mathcal{M} &=& \sum_{\lambda,\beta} M_\mathrm{met} (\lambda,\beta)   S(\lambda,\beta) \cos({\varphi}),\\
    \mathcal{E} &=& \sum_{\lambda,\beta,v_\mathrm{imp}} \frac{1}{2} M_\mathrm{met} (\lambda,\beta,v_\mathrm{imp}) v_\mathrm{imp}^2(\lambda,\beta) S(\lambda,\beta) \cos({\varphi}), \\
    \mathcal{P^+} &=& \mathcal{C}\sum_{\lambda,\beta,v_\mathrm{imp}} M_\mathrm{met} (\lambda,\beta,v_\mathrm{imp}) v_\mathrm{imp}^{2.46}(\lambda,\beta) S(\lambda,\beta) \cos^3({\varphi}), \label{EQ:EJECTA}
\end{eqnarray}
where $\lambda, \beta$ are the sun-centered ecliptic longitude and latitude of  meteoroids, $S(\lambda,\beta)$ is the shadow map coefficient, $\varphi$ is the incidence angle measured from the normal of the surface patch, \added{$v_\mathrm{imp}$ is the meteoroid impact velocity,} and $\mathcal{C} = 7.358 \mathrm{~km}^{-2} \mathrm{~s}^2$ is a scaling constant determined from the laboratory experiments reported by \citet{Koschny_Grun_2001}. \added{The scaling constant $\mathcal{C}$ describes the amount of ejecta the surface produces. The value of  $\mathcal{C}$ used here characterizes impacts of glass projectiles into ice-silicate surfaces. The mass of produced ejecta, $\mathcal{P}^+$, is likely orders of magnitude smaller than for solid ice surfaces, as suggested by the discrepancy of the meteoroid modeling and Lunar Dust Experiment discussed in \citet{Pokorny_etal_2019}. The incidence angle $\varphi$ is calculated for each surface patch separately, where $\cos{\varphi}=-\overrightarrow{n_\mathrm{sur}} \cdot \overrightarrow{e_\mathrm{imp}}$, where $\overrightarrow{n_\mathrm{sur}}$ is the normal of the surface patch, and $\overrightarrow{e_\mathrm{imp}}$ is the unit velocity vector of the impacting meteoroid. Then grazing impacts have $\cos{\varphi}$ close to zero, while perpendicular impacts are close to unity. }
Since we only  investigate the centroids (central points) of each surface triangular element, $S(\lambda,\beta)$ is either one or zero. We tested the shadowing procedure for more points (100) for each triangle for a smaller area (Shackleton crater) and the result did not yield significant differences when averaging over the entire year.
The ejecta mass production rate $\mathcal{P^+}$ in Eq. \ref{EQ:EJECTA} is a special case of a more general equation given by \citet{Koschny_Grun_2001}, which was applied in \citet{Pokorny_etal_2019} to compare to the ejecta cloud measured around the Moon by LDEX \citep[\textit{Lunar Dust EXperiment;}][]{Horanyi_etal_2015}. 

\section{Comparison of north and south polar regions}
\begin{figure}
\includegraphics[width=\textwidth]{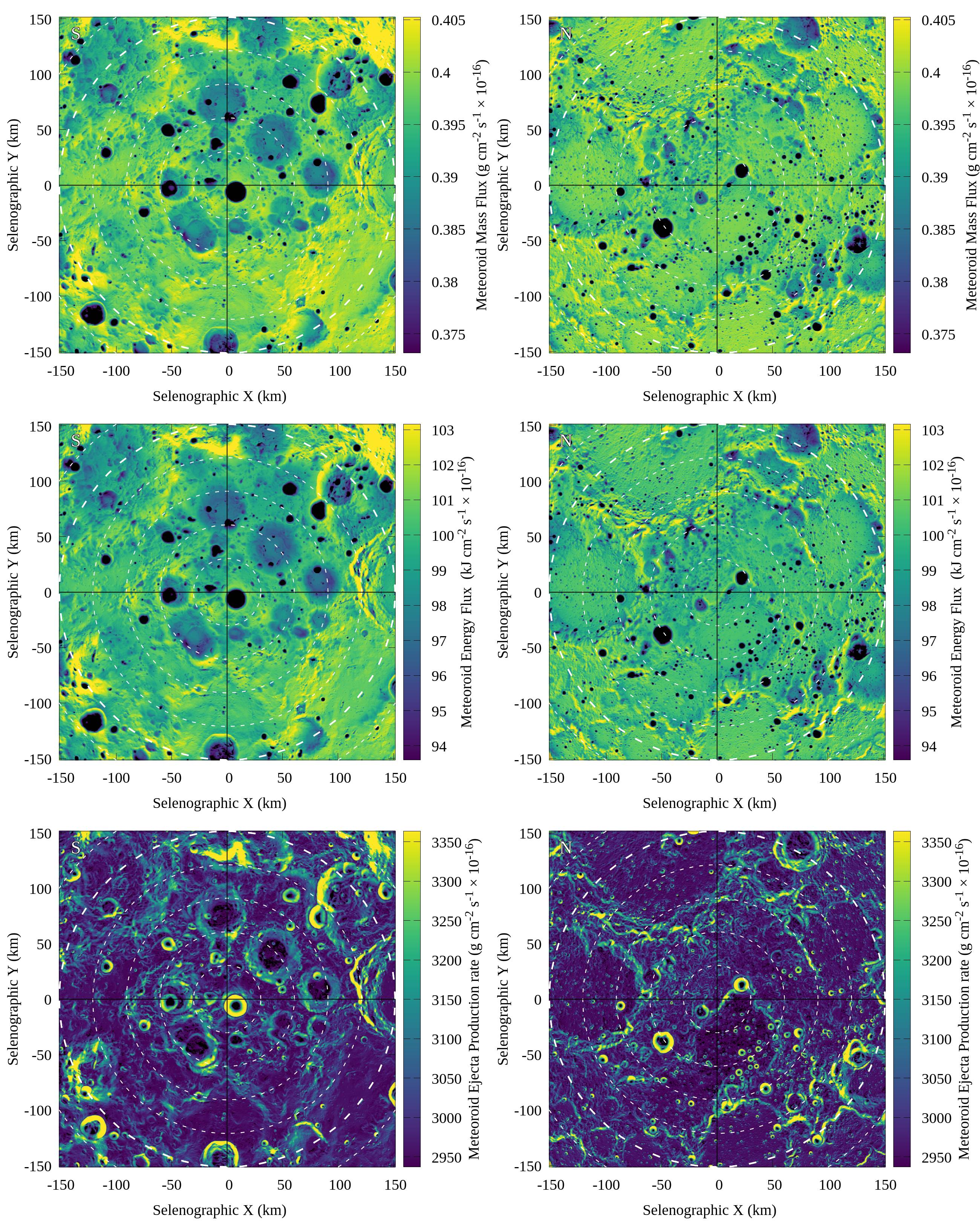}
\caption{\textit{Left panels:} South pole maps of the a) meteoroid mass flux $\mathcal{M}$, c) meteoroid energy flux $\mathcal{E}$ and e) meteoroid mass ejecta production rate $\mathcal{P^+}$. \textit{Right panels:} North pole maps of the a) meteoroid mass flux $M$, c) meteoroid energy flux $\mathcal{E}$ and e) meteoroid mass ejecta production rate $\mathcal{P^+}$. All maps are in gnomonic (selenographic) coordinates with resolution 250 m.}
\label{FIG:NORTH_SOUTH_ALL}
\end{figure}

In this Section we  present our results and compare the meteoroid environment at the two lunar polar regions. Figure \ref{FIG:NORTH_SOUTH_ALL} shows the variations of $\mathcal{M},\mathcal{E}$ and $\mathcal{P^+}$ for  the south (left panels) and north (right panels) polar regions. The color ranges in Fig. \ref{FIG:NORTH_SOUTH_ALL} do not show the entire range of values, but are limited to $[P_{2.275\%},P_{97.725\%}]$, where $P_x$ is the $x$-th percentile of the sample. 
Using percentile values allows us to show the variations of each quantity over the entire region, while avoiding regions with very low values (PSR) or very high values (mountains). We do not use the mean values and standard deviations here due to the non-normal distribution of  $\mathcal{M},\mathcal{E}$ and $\mathcal{P^+}$ on lunar poles. 

 Our model finds very small variations of the meteoroid mass flux, $\mathcal{M}$, at both poles, with median values equal to $\mathcal{M_\mathrm{S_{50\%}}}=0.3971 \mathrm{~g~cm}^2\mathrm{~s}^{-1}\times 10^{-16}$ and $\mathcal{M_\mathrm{N_{50\%}}}=0.3976 \mathrm{~g~cm}^2\mathrm{~s}^{-1}\times 10^{-16}$, where 95\% of values represented on the north and south polar maps  are within $\pm6\%$ of the median value (see Table \ref{TAB:ALL_DATA} for more details). The modeled minimum value near the south pole is $\mathcal{M}_\mathrm{min}=0.3077 \mathrm{~g~cm}^2\mathrm{~s}^{-1}\times 10^{-16}$  within a small $\sim$5 km crater between the Scott M and Nobile craters (Cartesian $[X=116, Y=130]$ km),  resulting in a 22\% smaller mass flux than the median value;  similar results are also found at the north pole. On the other hand, the maximum modeled value  at both poles is $\mathcal{M}_\mathrm{max}=0.4151 \mathrm{~g~cm}^2\mathrm{~s}^{-1}\times 10^{-16}$, which is only 5\%  higher than the median value. This means that despite lunar polar topography and the presence of PSRs,  the meteoroid mass flux is rather uniform across both polar regions, unlike the solar flux, which varies by orders of magnitude.

The modeled spatial variations of the meteoroid energy flux $\mathcal{E}$  are very similar to  those of the mass flux, where the only difference is higher energy fluxes on several ridges located between $4-5$ degrees away from the pole. The median values of $\mathcal{E}$ are very similar for both poles, $\mathcal{E_\mathrm{S_{50\%}}}=99.96$ and $\mathcal{E_\mathrm{N_{50\%}}}=100.11$ kJ cm$^{-2}$ s$^{-1}\times 10^{-16}$, respectively. Despite the complex topography of both lunar poles, from the variations of $\mathcal{M}$ and $\mathcal{E}$ we can estimate that the average impact velocity $V_\mathrm{imp}=22.4$ km s$^{-1}$ provides a reasonable evaluation of meteoroid energy from the meteoroid flux at both poles
This is in agreement with the results reported by \citet{Pokorny_etal_2019}, which showed in their Figure 9 that the longitudinally-averaged meteoroid impact velocity is between 15 and 23 km s$^{-1}$, and the majority of locations on both poles are not shielded from impactors coming from slightly above or below the ecliptic. Since energy scales with the square of the incident velocity, the more energetic impactors are emphasized, thus $V_\mathrm{imp}=22.4$ km s$^{-1}$ is higher than that inferred from meteoroid mass fluxes.

The increased dependence on the incidence angle $\varphi$ for the meteoroid ejecta production rate $\mathcal{P}^+$ changes the overall appearance of polar maps. While the majority of flat regions have very similar values close to the median $\mathcal{P}^+_\mathrm{S_{50\%}}=2990.45$ and $\mathcal{P}^+_\mathrm{N_{50\%}}=2973.18$ g cm$^{-2}$ s$^{-1}\times 10^{-16}$, all areas with higher slopes produce significantly more ejecta, which is valid even for permanently shadowed walls of craters on both lunar poles (e.g., the Shackleton crater close to the south pole). This is due to the fact, that the $\cos^3(\varphi)$ scaling for areas with higher slopes emphasizes the contribution of  energetic meteoroids originating from directions close to the ecliptic (the so-called apex source populated by Halley-type and Oort Cloud Comet meteoroids).  Unlike $\mathcal{M}$ or $\mathcal{E}$, Table \ref{TAB:ALL_DATA} also shows that $\mathcal{P}^+$ is highly asymmetric for both poles, where the first 50\% of data points are concentrated very close to the median, while the regions with higher $\mathcal{P}^+$ values exhibit a long tail.

\begin{table}[]
\begin{center}
\begin{tabular}{l|rrrrrrr}
& \multicolumn{7}{c}{Percentile}                                                                                                                                                                  \\ \cline{2-8} 
                      & \multicolumn{1}{c}{0.135\%} & \multicolumn{1}{c}{2.275\%} & \multicolumn{1}{c}{15.73\%} & \multicolumn{1}{c}{50\%} & \multicolumn{1}{c}{84.27\%} & \multicolumn{1}{c}{97.725\%} & \multicolumn{1}{c}{99.865} \\ \hline \hline
North $\mathcal{M}$   & 0.3378                    & 0.3740                    & 0.3913                    & 0.3976                 & 0.4002                    & 0.4031                     & 0.4063                     \\
South $\mathcal{M}$   & 0.3310                    & 0.3732                    & 0.3898                    & 0.3971                 & 0.4009                    & 0.4052                     & 0.4128                     \\
North $\mathcal{E}$ &     82.027 &   93.832 &   98.444 &  100.11 &  100.95 &  102.32 &  103.95 \\
South $\mathcal{E}$ &     79.842 &   93.594 &   97.991 &   99.956 &  101.21 &  103.16 &  106.92 \\
North $\mathcal{P}^+$ & 2908.8 & 2936.7 & 2944.7 & 2973.2 & 3089.8 & 3311.2 & 3524.4 \\
South $\mathcal{P}^+$ & 2904.6 & 2937.0 & 2947.9 & 2990.4 & 3119.9 & 3363.6 & 3596.5 \\               
\end{tabular}
\caption{Percentile values for $\mathcal{M}$ (g cm$^{-2}$ s$^{-1}\times10^{-16}$), $\mathcal{E}$ (kJ cm$^{-2}$ s$^{-1}\times10^{-16})\mathrm{~and~} \mathcal{P}^+$ (g cm$^{-2}$ s$^{-1}\times10^{-16}$) for both the north and south lunar poles. \label{TAB:ALL_DATA}}
\end{center}
\end{table}

The Shackleton crater is a good example for understanding this effect. The minimum value of $\mathcal{P}^+=2743.14$ g cm$^{-2}$ s$^{-1}\times 10^{-16}$  is localized at the floor of the crater, while the sides of the crater with high slopes \citep[up to 32$^\circ$, ][]{Zuber_etal_2012} experience 31\% higher values of $\mathcal{P}^+=3605.42$ g cm$^{-2}$ s$^{-1}\times 10^{-16}$. When comparing it to values in Table \ref{TAB:ALL_DATA}, we see that Shackleton crater experiences both extremes of the meteoroid ejecta production rate. Nevertheless, the floor of the crater is still subjected to meteoroid bombardment, despite the very high angles of incidence ($28-32^\circ$, i.e., effectively shielding the crater floor from meteoroids with ecliptic latitudes $<32^\circ$).

\begin{figure}
\includegraphics[width=\textwidth]{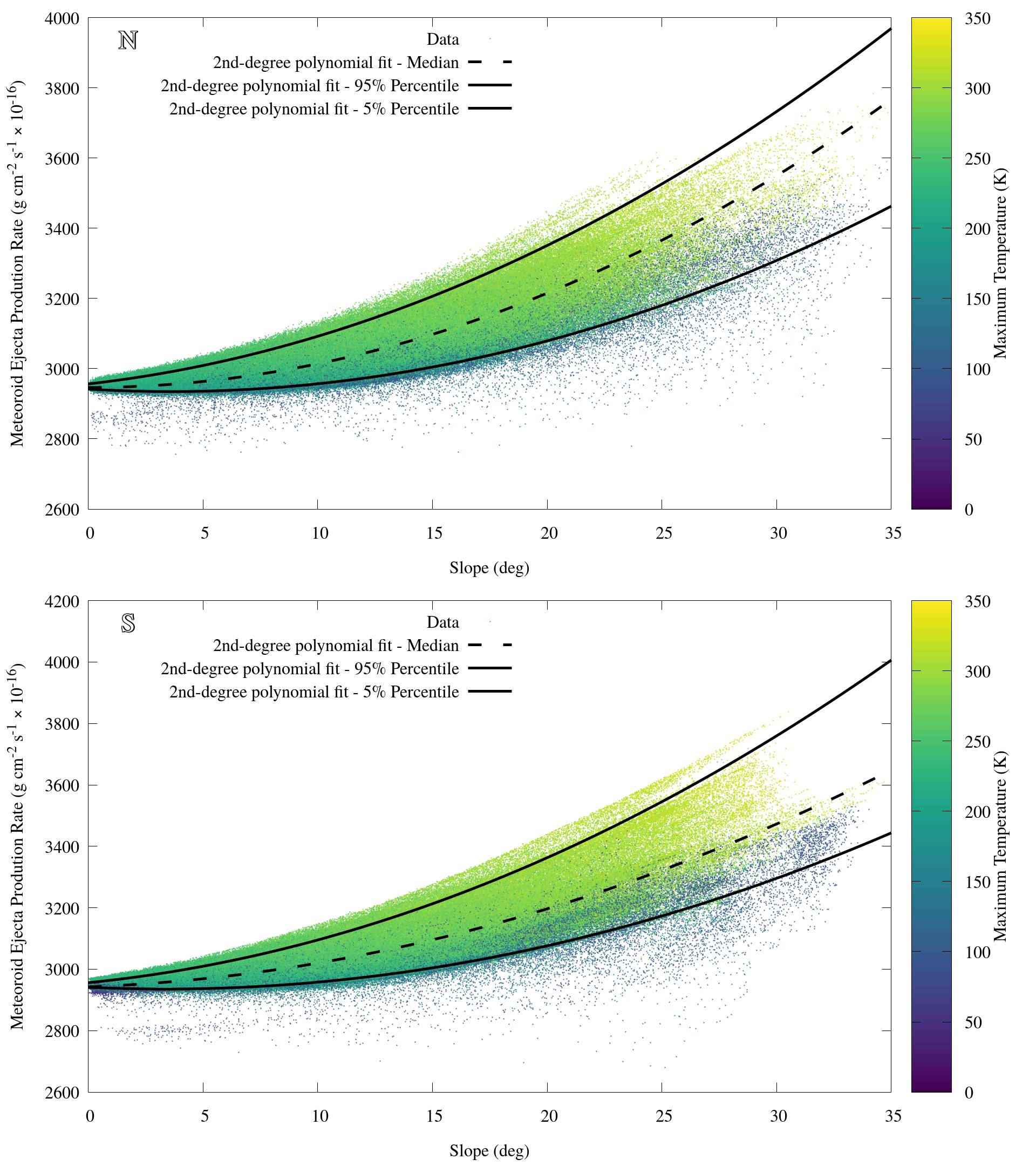}
\caption{\textit{Top panel:} Meteoroid ejecta mass production rates $\mathcal{P^+}$ as a function of the surface slope in degrees \added{for the lunar north pole}. The color bar show the maximum surface temperature $T_\mathrm{max}$ in Kelvin. The ejecta mass production rate follows a quadratic trend with respect to the surface slope shown by the dashed black line (quadratic fit to the median value for each $0.5^\circ$ bin) and solid black lines (quadratic fit to the 90\% percentile range). \textit{Bottom panel}: The same as the top panel but for the lunar \replaced{north}{south} pole.}
\label{FIG:SLOPE_EJECTA}
\end{figure}

 Figure \ref{FIG:SLOPE_EJECTA}  shows how the meteoroid ejecta production rate $\mathcal{P}^+$ varies with slope on the global scale. Each data point in this figure represents the value modeled at each surface patch on our maps, where the color coding is based on the maximum temperature registered at the particular location of the data point. In order to assess the trend in the data, we calculated three percentile values (5\%, 50\%, and 95\%) using a moving window in slope with a width of $1^\circ$, taking steps of $0.5^\circ$ in the range $0^\circ-20^\circ$, and then fitted the percentile ranges with 0-5 degree polynomials. We find that polynomials  of degree higher than 2nd-degree do not significantly improve the fit \added{(using a reduced $\chi^2$ statistic with unit variance $\sigma=1$} , thus we opted for the 2nd-degree polynomial as our fitting function for the simplicity  of its parameters. Both poles show a very similar trend where the meteoroid ejecta production rate $\mathcal{P}^+$ for all three percentile values is well described with the 2nd-degree polynomial. However, the south pole, due to its more diverse topography and larger permanently shadowed area, shows a larger spread of $\mathcal{P}^+$ values, but the quadratic trend holds for both the low and high maximum temperature areas. The north pole has a very similar envelope ($5\%-95\%$ percentile range) to the south pole, but its median value grows more rapidly with increasing slope, which is due to the smaller amount of highly-inclined permanently shadowed areas. All coefficients for the 2nd-degree polynomial fits are summarized in Table \ref{TAB:QUAD_FITS}. \added{These polynomial fits can be easily used for a quick estimate of the ejecta production rates in any model that requires such an input, e.g. a model estimating impact gardening on lunar poles. The polynomial fits of the median and the $5\%-95\%$ percentile range allow quite precise estimates without using the full-fledged model presented here.}

\begin{figure}
\includegraphics[width=\textwidth]{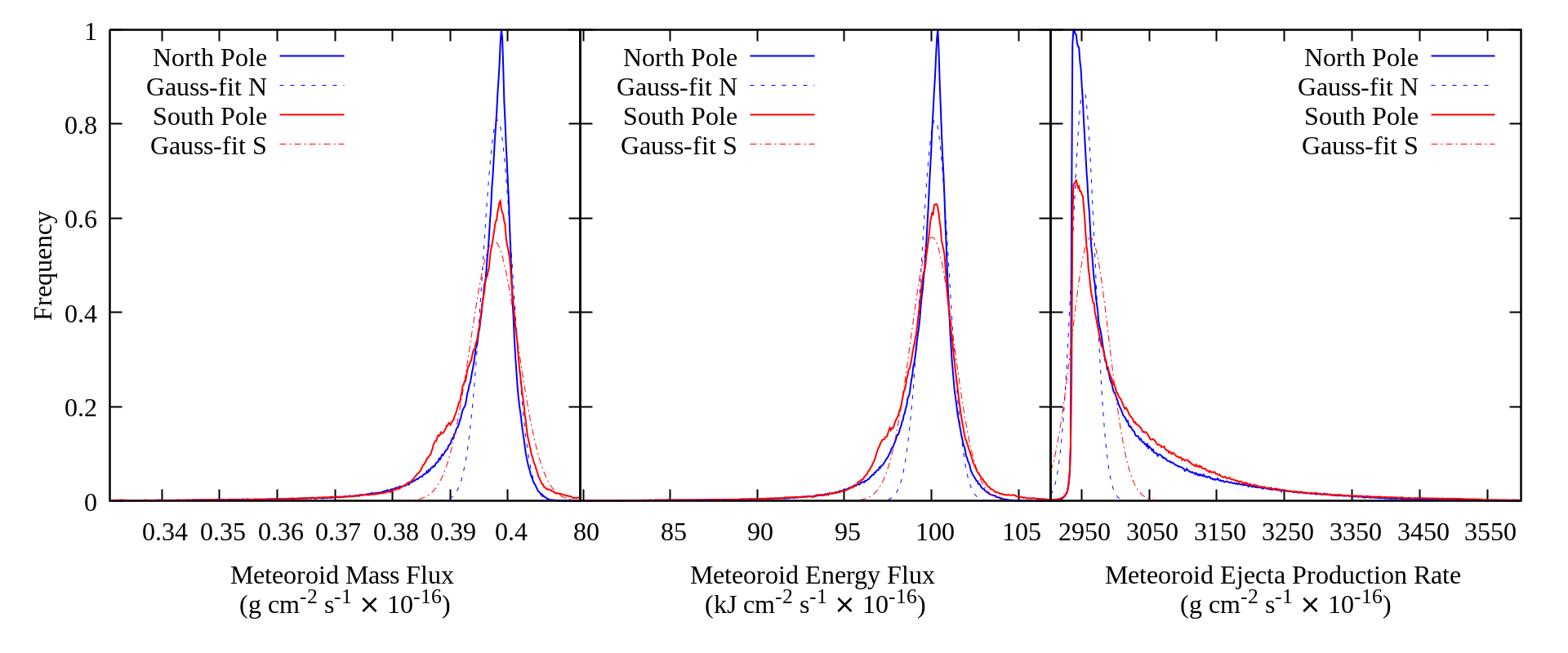}
\caption{Frequencies of occurrences for $\mathcal{M}$ (left panel), $\mathcal{E}$ (middle panel) and $\mathcal{P}^+$ (right panel) normalized to the maximum frequency on either poles. \added{Blue solid lines indicate values for the north pole, whereas blue dashed lines are Gaussian fits to their respective quantities. In red are shown their south pole equivalents.}}
\label{FIG:HISTOGRAMS_ALL}
\end{figure}

Finally, we may look at the histograms of modeled $\mathcal{M}, \mathcal{E}$ and $\mathcal{P}^+$ to demonstrate the divergence from the normal (Gaussian) distribution that is usually assumed for various physical processes. Figure \ref{FIG:HISTOGRAMS_ALL} shows that the histograms of meteoroid mass and energy fluxes are asymmetric, with a tail toward the smaller values where the function profiles show a shape similar to a Cauchy-Lorentz distribution. Frequencies of both quantities are more dispersed for the south pole due to the more complex topography. The distribution of meteoroid ejecta production rate $\mathcal{P}^+$ (right panel in Fig. \ref{FIG:HISTOGRAMS_ALL}) is extremely asymmetric, where the majority of locations on both poles have similar values of $\mathcal{P}^+$, which is also apparent from Figs. \ref{FIG:NORTH_SOUTH_ALL} and \ref{FIG:SLOPE_EJECTA} where the low slope ($<5^\circ$) areas are experiencing essentially the same effects of the meteoroid bombardment. \added{To show the divergence of these distribution from normal distributions, we added the best normal distribution fits (using the least-squares method) to $\mathcal{M}, \mathcal{E}$ and $\mathcal{P}^+$ on both poles (dashed lines in Fig. \ref{FIG:HISTOGRAMS_ALL}}).

\begin{table}[]
\begin{center}

\begin{tabular}{r||ccc|ccc}
\multicolumn{1}{l||}{} & \multicolumn{3}{c|}{South} & \multicolumn{3}{c}{North} \\ \cline{2-7} 
\multicolumn{1}{l||}{} & a       & b       & c      & a       & b       & c      \\ \hline \hline
5\% Percentile        & 0.5078  & -3.4082 & 2941.2 & 0.5350  & -3.8322 & 2941.7 \\
50\% Percentile       & 0.4978  & 2.7356  & 2942.7 & 0.6682  & 0.0638  & 2946.4 \\
95\% Percentile       & 0.6443  & 7.4543  & 2956.2 & 0.6147  & 7.4240  & 2956.5
\end{tabular}
\caption{Coefficients for quadratic function fits ($f(x)=ax^2+bx+c$) of the meteoroid ejecta production rate vs. slope dependency for both north and south lunar poles. \label{TAB:QUAD_FITS}}
\end{center}
\end{table}

The contributions of different meteoroid source populations to impact related processes on lunar poles are summarized in Table \ref{TAB:DIFF_SOURCES}. The median values calculated for a combination of both poles show that JFC meteoroids dominate the meteoroid mass flux $\mathcal{M}$ by providing 65\% of the total sum. However, this percentage changes for both the meteoroid energy flux $\mathcal{E}$ where HTC meteoroids lead with 35\%, followed by OCC meteoroids with 33\% and JFC meteoroids with 31\%. This change is due to the higher impact velocities of HTC and OCC meteoroids. $\mathcal{P}^+$ follows a similar trend with HTC meteoroids aquiring almost 45\% of the total sum due to their larger spread in ecliptic latitudes and thus easier access of both lunar poles. With the exception of the meteoroid mass flux where MBA meteoroids represent 7\% of the total mass flux, the energy flux and ejecta production rates are close to 1\% and thus the MBA contribution can be neglected.

The meteoroid environment bombarding the moon is calibrated using the terrestrial mass flux with the following values: $\mathrm{MBA}=3,700, \mathrm{JFC}=34,600,\mathrm{HTC}=2,820,\mathrm{OCC}=2,180$ \deleted{t d$^{-1}$} \added{kg per day} \citep{CarrilloSanchez_etal_2016,Pokorny_etal_2019}. In order to better quantify the effects of the meteoroid impact geometry and lunar topography, we normalized the terrestrial mass flux for each meteoroid source to \deleted{1 t d$^{-1}$} \added{1,000 kg per day} (bottom part of Table \ref{TAB:DIFF_SOURCES}). The normalized mass flux is very similar for MBA and JFC meteoroids (12-13\%) but the long-period comet sources are much more efficient in delivering mass to the lunar poles since their ecliptic latitudes spread further from the ecliptic. Furthermore, due to their higher impact velocities, the long-period sources dominate both $\mathcal{E}$ and $\mathcal{P}^+$ with $>95\%$ of the total from all sources. Only due to their abundance in the inner solar system, JFC meteoroids are a significant contributor to the space weathering processes on lunar poles.

\begin{table}[]
\begin{center}
\begin{tabular}{l|p{0.1\textwidth}p{0.1\textwidth}p{0.1\textwidth}p{0.1\textwidth}}
                & \multicolumn{1}{l}{MBA} & \multicolumn{1}{l}{JFC} & \multicolumn{1}{l}{HTC} & \multicolumn{1}{l}{OCC} \\ \cline{2-5} 
                & \multicolumn{4}{c}{Median Value}                                                                     \\ \hline \hline
$\mathcal{M}$   & 0.0264                  & 0.2595                  & 0.0682                  & 0.0431                  \\
$\mathcal{E}$   & 1.1618                  & 30.724                  & 35.325                  & 32.700                 \\
$\mathcal{P}^+$ & 33.476                  & 737.88                  & 1334.6                  & 891.94                \\
                & \multicolumn{4}{c}{Percentage (\%)}                                                                       \\ \hline\hline
$\mathcal{M}$   & 6.6465                  & 65.332                 & 17.170                 & 10.851                 \\
$\mathcal{E}$   & 1.1628                  & 30.751                 & 35.357                 & 32.729                 \\
$\mathcal{P}^+$ & 1.1166                  & 24.614                 & 44.517                 & 29.753                 \\
                & \multicolumn{4}{c}{Normalized to {1,000 kg per day} at Earth (\%)}                                   \\ \hline\hline
$\mathcal{M}$   & 12.178                  & 12.801                  & 41.277                 & 33.744                 \\
$\mathcal{E}$   & 1.0932                  & 3.0909                  & 43.603                 & 52.213                 \\
$\mathcal{P}^+$ & 0.9912                  & 2.3364                  & 51.847                 & 44.825                
\end{tabular}
\caption{\textit{Top:} Median values for $\mathcal{M}$ (g cm$^{-2}$ s$^{-1}\times10^{-16}$), $\mathcal{E}$ (kJ cm$^{-2}$ s$^{-1}\times10^{-16})\mathrm{~and~} \mathcal{P}^+$ (g cm$^{-2}$ s$^{-1}\times10^{-16}$) for both lunar poles differentiated by four meteoroid source populations. \textit{Middle:} Percentages for each population. \textit{Bottom:} Percentages for each population normalized to \deleted{1 t d$^{-1}$} \added{1,000 kg per day}for each source population. The calibrated mass flux at Earth used here adopted from \deleted{\citet{Pokorny_etal_2019} is $\mathrm{MBA}=3.7, \mathrm{JFC}=34.6,\mathrm{HTC}=2.82,\mathrm{OCC}=2.18$ t d$^{-1}$}\added{$\mathrm{MBA}=3,700, \mathrm{JFC}=34,600,\mathrm{HTC}=2,820,\mathrm{OCC}=2,180$ kg per day}.  \label{TAB:DIFF_SOURCES}}
\end{center}
\end{table}

How is it possible to achieve such uniform distributions of the mass flux, energy flux or ejecta production rate over such a variable topography? \citet{Pokorny_etal_2019} showed that the meteoroids impacting the Moon have a very broad distribution of the ecliptic latitude (i.e., the angular distance measured from the ecliptic in the $z$-direction), thus even deep polar craters are exposed to a considerable meteoroid flux, especially the north/south toroidal source observed at Earth \citep{CampbellBrown_2008, Janches_etal_2015} originating from long-period comets \citep{Nesvorny_etal_2011OCC,Pokorny_etal_2014}. Moreover, the abundant and high energy meteoroid sources close to the ecliptic impact the lunar poles at very shallow angles, and due to the cosine dependence, their contribution will be considerably diminished. This effect is even more accentuated for the ejecta mass production rates $\mathcal{P^+}$ due to the steeper dependency on the incidence angle \citep[see e.g.][]{Gault_1973} and thus the locations with small slopes experience negligible contributions from low inclination meteoroids.

\section{Permanently shadowed regions with low slopes - Regions without mass wasting}
In this Section we focus on regions that should be unaffected by mass wasting and should be able to sustain water-ice deposits for geological timescales.
For such regions we choose the surface slope $\alpha<10^\circ$ based on work of \citet{Lucey_etal_2014}, who concluded that $\alpha<10^\circ$ is safe from mass wasting. Furthermore we analyze only regions where the maximum surface temperature $T_\mathrm{max}<110$ K, i.e., regions that should be able to sustain ice deposits for geological timescales. \added{Independent remote spacecraft sensing data from the Moon Mineralogy Mapper \citep{Li_etal_2018} and LRO \citep{FISHER_ETAL_2017} show that many of these regions should indeed harbor deposits of water-ice}.

The motivation for a more detailed analysis of such regions stems from the recent works regarding space weathering and surface regolith alteration in and around south pole craters that are partially permanently shadowed and partially exposed to solar irradiation and solar wind \citep{BYRON_ETAL_2019}. Such places, that are unique only to polar regions, are fortuitous laboratories allowing us to study different processes of lunar soil alteration separately. Mass wasting on more inclined slopes exposes the fresh, not weathered material \citep[e.g.,][]{Zuber_etal_2012}, which effectively restarts the space weathering process and adds complexity into the soil alteration analysis.

\begin{figure}
\includegraphics[width=\textwidth]{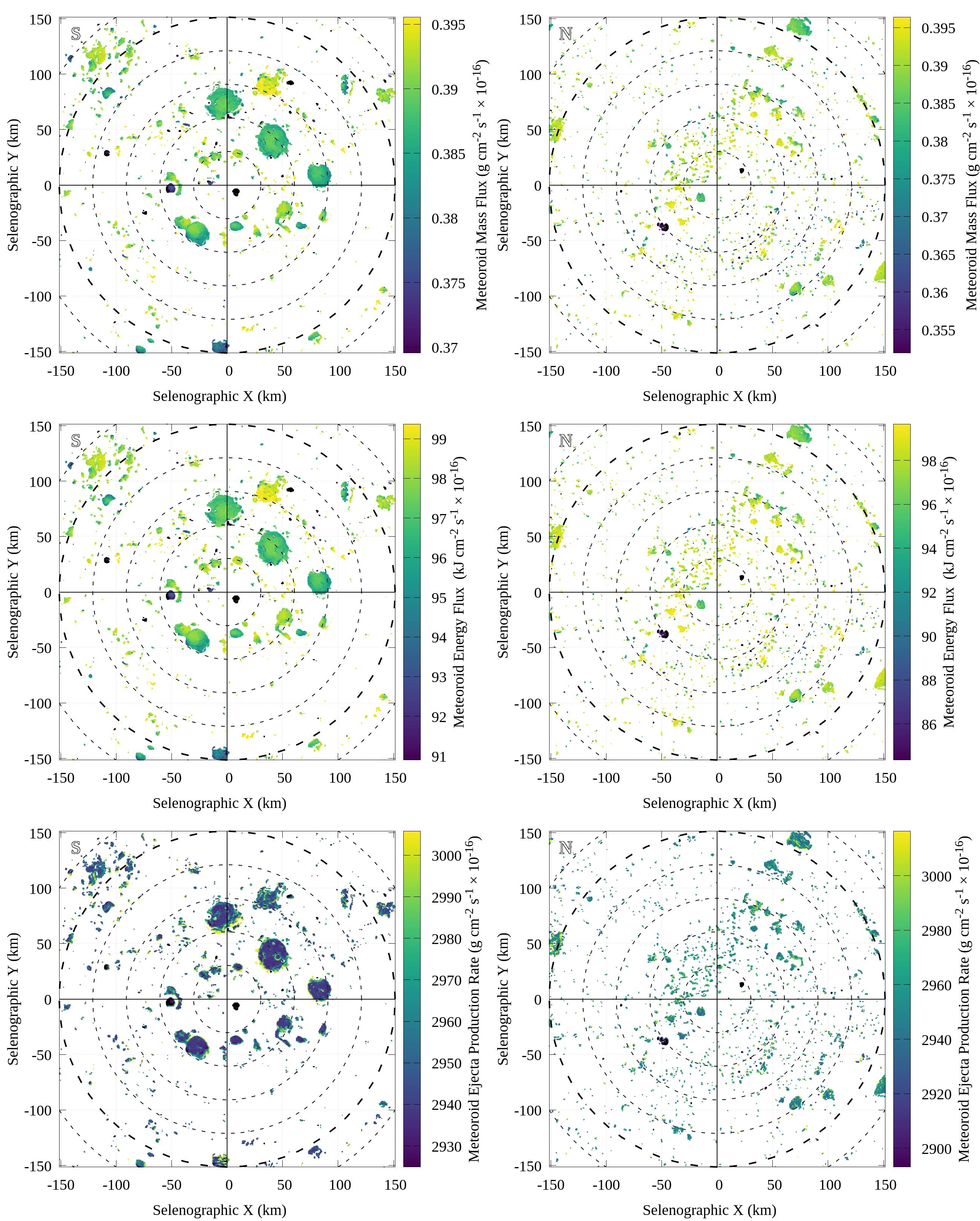}
\caption{The same as Fig. \ref{FIG:NORTH_SOUTH_ALL} but now only for low slope ($<10^\circ$) water ice stable ($T_\mathrm{max}<110$ K) regions.}
\label{FIG:NORTH_SOUTH_FILTERED}
\end{figure}

\begin{figure}
\includegraphics[width=\textwidth]{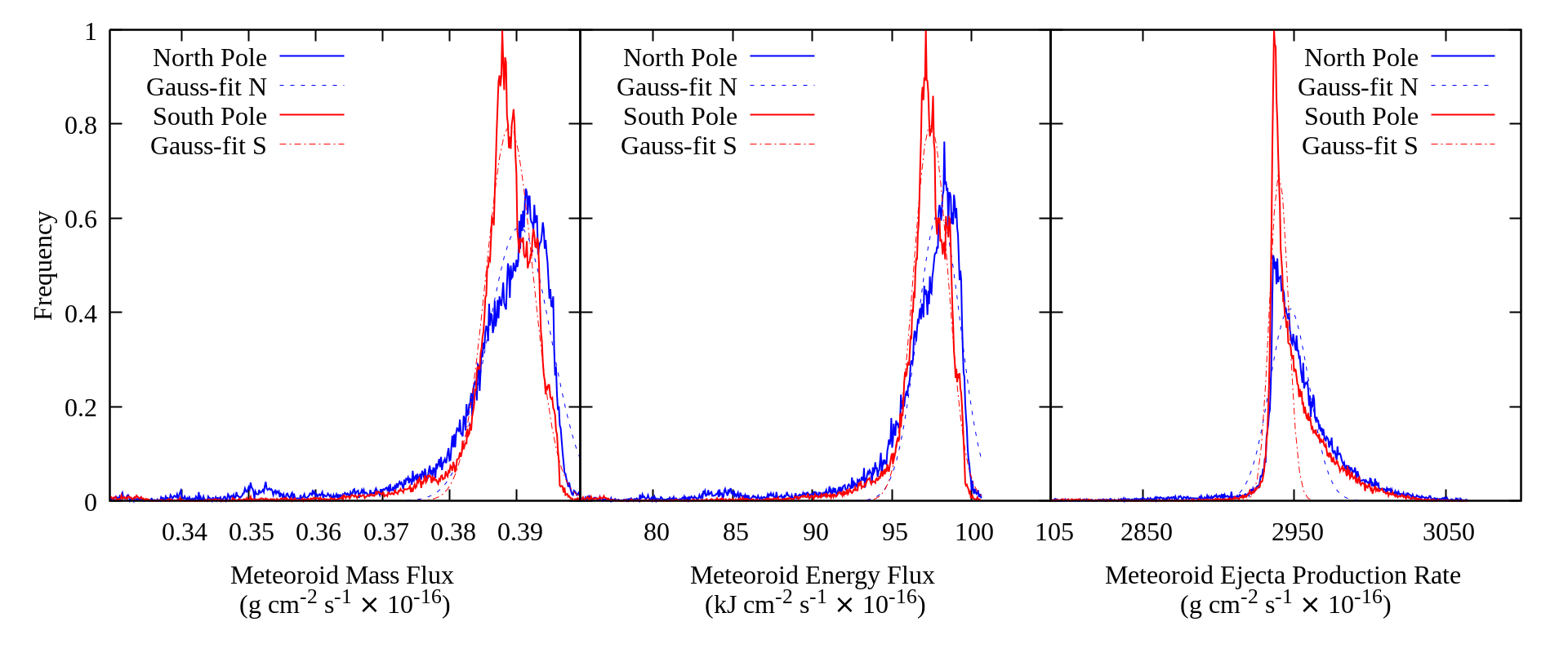}
\caption{The same as Fig. \ref{FIG:HISTOGRAMS_ALL} but now only for low slope ($<10^\circ$) water ice stable ($T_\mathrm{max}<110$ K) regions.}
\label{FIG:HISTOGRAMS_FILTER}
\end{figure}

Figure \ref{FIG:NORTH_SOUTH_FILTERED} shows  the variations of $\mathcal{M}, \mathcal{E}$ and $\mathcal{P}^+$ for all regions that satisfy our conditions. We see that the topography of the north and south poles is very different, where in the south the flat cold and stable areas (CSAs) are mostly concentrated in several craters, while at the north pole CSAs are more scattered and also less abundant (approximately 4500 km$^2$ vs 2130 km$^2$ for south and north respectively). Despite significant differences in the CSA distribution at both locations, the mass flux distribution is very similar for both lunar poles, which is more clearly shown in the frequency histograms (Fig. \ref{FIG:HISTOGRAMS_FILTER}). When compared to Fig. \ref{FIG:HISTOGRAMS_ALL}, the distributions are shifted toward the lower values due to being shadowed from the low inclination meteoroid impacts, however the median values are only 2\% smaller than the median values calculated for the entire map. Only craters with high slope angles, $\sim30^\circ$ like Shackleton, are shielded more efficiently, however neither the mass flux nor the energy flux  decrease by more than 20\% below the median values.

The meteoroid ejecta production rate exhibits somewhat more variation, but within different craters it does not reach more than a few percent. We conclude that the ice deposits should not exhibit significantly different aging signatures due to the meteoroid impacts. Perhaps the more important effect is the meteoroid-induced mass wasting from the crater sides with higher slopes.  In these places the rims are being excavated significantly more efficiently than the floor, and the material excavated from the crater sides may be flowing into the crater and burying the top layers of its floor. \added{Ejecta trajectories and model mass wasting caused by meteoroids might be investigated in future work.} 

\section{Discussion}
The meteoroid dynamical model used here \citep{Pokorny_etal_2019} is a product of several years of work that combines the dynamical evolution of meteoroid-generating populations in the solar system (main-belt asteroids, short/long-period comets) and various observed constraints. The meteoroid model has inherent uncertainties due to the complex nature of both the dynamical modeling and its intrinsic free parameters, as well as the uniqueness of each constraint (meteor orbital distribution at Earth, shape of the Zodiacal cloud, spacecraft measured particle flux). \citet{Pokorny_etal_2019} quantified the effect of several free parameters in their model, showing it to be quite robust to such uncertainty. \added{The three most influential free parameters in their models were found to be: (a) the collisional lifetime multiplier, $F_\mathrm{coll}$, that effectively scales the flux of meteoroids with diameters $D>200~\mu$m \citep[see Fig. 6 in][]{Pokorny_etal_2019}; (b) the differential size-frequency index $\alpha$ that scales the relative abundance of meteoroids ejected from their sources; and (c) the mixing of meteoroid populations at Earth. The effects of different combinations of (a) $F_\mathrm{coll}\in[10,50]$ and (b) $\alpha \in [3.4,4.6]$ was $\pm10\%$ with respect to the overall mass flux on the Moon, and about $\pm17\%$ in terms of the energy flux. These variations were with respect to the entire Moon surface. When only the polar regions are taken into account, the maximum divergence from the model solution used in this study is $<10\%$ for all three main quantities considered here: $\mathcal{M}, \mathcal{E}\mathrm{~and~}\mathcal{P}^+$.}

The population mixing (i.e. the ratio between asteroidal and cometary sources) adopted here relies strongly on the work reported by \citet{CarrilloSanchez_etal_2016} and \textit{LDEX} measurements that \citet{Pokorny_etal_2019} used to constrain the ratio of Halley-type and Oort Cloud Comets. Here, we present the results as the currently best dynamical model we can provide. Should a new set of constraints emerge that would significantly change the quantities investigated here, the model could be updated in the future.

 Even though  meteoroid gardening might be the dominant source of surface water-ice removal on lunar poles \citep{HAYNE_ETAL_2015}, there is no significant difference in the distribution of the meteoroid flux or meteoroid ejecta production rate between the two lunar poles based on our calculations. One factor that might change the absolute values of meteoroid ejecta production rates presented here by orders of magnitude are the assumed ejecta yields. Due to the lack of experiments at speeds  $V_\mathrm{imp}\gtrsim 10$ km s$^{-1}$ , we can only speculate on the effects of higher impact velocities to the ejecta rate from different surface materials. However, the comparison of the meteoroid impact generated ejecta to the lunar dust cloud density in \citet{Pokorny_etal_2019} suggests that the regolith ejecta yield might be orders of magnitude smaller than that of ice-silicate surfaces used in \citet{Koschny_Grun_2001}. This effect  could potentially have major consequences on excavation of pure ice deposits in PSRs, because pure ice or surface-exposed ice would be excavated on shorter time scales than ice-regolith mixtures or pure regolith layers. Thus the differences in ejecta yields between different materials might be more important than differences in meteoroid access to these regions. Although beyond the scope of this manuscript, it should be possible to quantify the erosion of different regions showing water-ice signatures using the calculations presented here.


\citet{HARUYAMA_ETAL_2013} tackled the problem of the water-ice presence inside the Shackleton crater by analyzing the \textit{SELENE (Selenological Engineering Explorer)} mission observations and concluded that the high reflectance regions inside the crater are due to the presence of pure anorthosite. \deleted{This conclusion differs from \textit{LRO LOLA} observations that suggest more than 20\% of water-ice \citep{Zuber_etal_2012}.}\added{\citet{Zuber_etal_2012} provided an alternative explanation that the \textit{LRO LOLA} reflectance could be explained by 20\% water-ice content.} In Fig. \ref{FIG:NORTH_SOUTH_ALL} we showed that the internal part of the Shackleton crater is subject to higher meteoroid ejecta production rates because  high slopes are more susceptible to high energy impacts, which suggests increased mass wasting on Shackleton's inner walls. Such impacts  should excavate fresh material that might be transported deeper into the crater. This idea is also supported by the depletion in craters on both Shackleton's rim and floor \citep{TYE_ETAL_2015}, which is most likely caused by extensive mass wasting induced by meteoroid impacts.

\added{Even though the impact driven excavation of water-ice is a complex mechanism and its modeling is beyond the scope of this manuscript, initial estimates can be made from the quantities presented in Figure \ref{FIG:NORTH_SOUTH_ALL}. The meteoroid energy flux $\mathcal{E}=100 \times 10^{-16}$ kJ cm$^{-2}$ s$^{-1}$ can be converted to the impact vaporization rate, $\mathcal{V}=6 \times 10^{-16}$ g cm$^{-2}$ s$^{-1}$, using iron projectiles, a temperature of 400K, and the quadratic term from Eq. 10 in \citet{Cintala_1992}. Assuming the regolith bulk density $\rho=1.5$ g cm$^{-3}$, the impacts vaporize $4.5 \times 10^{-16}$ mm per second, which translates to $1.3 \times 10^{-7}$ mm per year. \citet{Morgan_Shemansky_1991} state that the local interstellar medium H Ly-$\alpha$ radiation is the most efficient destruction process of water-ice which amounts to $7 \times 10^{-8}$ mm per year, i.e an effect comparable in magnitude to meteoroid bombardment. Both effects are not significantly affected by the polar topography, thus a proper modeling of impact gardening is needed to fully cover all subtle effects that regulate the water-ice stability on lunar poles.}

\section{Conclusions}
We present the first quantification of the meteoroid mass flux, meteoroid energy flux, and meteoroid ejecta production rate based on dynamical models and reflecting the topography of both lunar poles. The main conclusions can be summarized as:
\begin{itemize}

    \item Despite the complex topography of lunar poles and the presence of many permanently shadowed regions, the meteoroid-induced processes are rather uniform, with maximum deviations of 30\% from the median values of these quantities;
    \item Unlike the solar wind, meteoroid impacts can provide the energy needed for surface weathering (e.g., formation of nanophase iron, band-depth reduction) even at the deepest polar craters; 
    \item The local meteoroid ejecta production rate can be expressed as a 2nd-degree polynomial function of the surface slope, where the higher surface slopes are subject to higher excavation rates by meteoroids from long-period comets. This result promotes substantial mass wasting;
    \item Crater walls with high slopes such as Shackleton are being constantly reprocessed by the meteoroid bombardment, which exposes  fresher material. Additionally, as a result of the efficient mass wasting, the material on the floor of such craters will be buried by both the constant meteoroid mass flux and the material sliding from the crater walls;
    \item Due to small variations of meteoroid bombardment effects, we suggest that different excavation rates of ice deposits are due to different ejecta yields of various mixtures of ice-regolith surface layers.
\end{itemize}

The relative uniformity of meteoroid-induced processes from crater to crater results from the finding that the meteoroid environment of the Moon is composed of meteoroids with a broad range of inclinations which reduces the shadowing effects. If more pronounced differences across craters are attributable to meteoroid impacts than shown here, such a finding could imply higher relative importance of long-period comets presently or in the past. Such differences might also arise from significant differences in ejecta yields for different materials (e.g., ice-free soils vs ice).

All quantities calculated in this work for both lunar poles are available at \url{https://github.com/McFly007/AstroWorks/tree/master/Pokorny_et_al_2019_ApJ} and cover $300$ km $\times~300$ km area centered at both poles, i.e. approximately 5 degrees away from the pole. We combined our data set with the DIVINER output for convenience, so the values of the maximum temperature, average temperature, albedo and depth of ice are readily available. The complete description of the data products can be found in the project directory together with future updates. 

\acknowledgements
PP, DJ and MS's work was supported with NASA's SSO, LDAP and ISFM awards. The data products, codes and data used to generate plots in this manuscript can be found at: \url{https://github.com/McFly007/AstroWorks/tree/master/Pokorny_et_al_2019_ApJ}.

\appendix
\section{Data set overview}
\label{Sec:AppendixA}
\subsubsection*{Topography}
\noindent Reference for these data sets is \citet{Smith_etal_2017}
\noindent LRO LOLA GDR in polar sterographic projection for the south lunar pole (240m/pixel) resolution:\\
\noindent {\small\url{http://imbrium.mit.edu/DATA/LOLA_GDR/POLAR/FLOAT_IMG/LDEM_75S_240M_FLOAT.IMG}}
\\

\noindent LRO LOLA GDR in polar sterographic projection for the north lunar pole (240m/pixel) resolution:\\ \noindent {\small\url{http://imbrium.mit.edu/DATA/LOLA_GDR/POLAR/FLOAT_IMG/LDEM_75N_240M_FLOAT.IMG}}
\\

\subsubsection*{Average temperatures, maximum temperatures, depth to water ice permafrost}
\noindent Reference for these data sets is \citet{Paige_etal_2010}
\noindent LRO Diviner Lunar Radiometer Polar Resource Products - south lunar pole:\\
\url{http://pds-geosciences.wustl.edu/lro/lro-l-dlre-4-rdr-v1/lrodlr_1001/data/prp/dlre_prp_south.tab}\\

\noindent LRO Diviner Lunar Radiometer Polar Resource Products - north lunar pole:\\
\url{http://pds-geosciences.wustl.edu/lro/lro-l-dlre-4-rdr-v1/lrodlr_1001/data/prp/dlre_prp_north.tab}

\subsubsection*{Albedo}
\noindent Reference for these data sets is \citet{Lemelin_etal_2016}
\noindent LRO LOLA surface albedo maps in polar stereographic projection for the south lunar pole (1000m/pixel) resolution:\\
\url{http://imbrium.mit.edu/DATA/LOLA_GDR/POLAR/FLOAT_IMG/LDAM_50S_1000M_FLOAT.IMG} \\

\noindent LRO LOLA surface albedo maps in polar stereographic projection for the north lunar pole (1000m/pixel) resolution:\\
\url{http://imbrium.mit.edu/DATA/LOLA_GDR/POLAR/FLOAT_IMG/LDAM_50N_1000M_FLOAT.IMG}
\bibliography{papers}

\begin{thebibliography}{}
\expandafter\ifx\csname natexlab\endcsname\relax\def\natexlab#1{#1}\fi
\providecommand{\url}[1]{\href{#1}{#1}}

\bibitem[{{Arnold}(1975)}]{Arnold_1975}
{Arnold}, J.~R. 1975, Lunar and Planetary Science Conference Proceedings, 2,
  2375

\bibitem[{{Byron} {et~al.}(2019){Byron}, {Retherford}, {Greathouse}, {Mand t},
  {Hendrix}, {Poston}, {Liu}, {Cahill}, \& {Mazarico}}]{BYRON_ETAL_2019}
{Byron}, B.~D., {Retherford}, K.~D., {Greathouse}, T.~K., {et~al.} 2019,
  Journal of Geophysical Research (Planets), 124, 823

\bibitem[{{Campbell-Brown}(2008)}]{CampbellBrown_2008}
{Campbell-Brown}, M.~D. 2008, \icarus, 196, 144

\bibitem[{{Carrillo-S{\'a}nchez} {et~al.}(2016){Carrillo-S{\'a}nchez},
  {Nesvorn{\'y}}, {Pokorn{\'y}}, {Janches}, \&
  {Plane}}]{CarrilloSanchez_etal_2016}
{Carrillo-S{\'a}nchez}, J.~D., {Nesvorn{\'y}}, D., {Pokorn{\'y}}, P.,
  {Janches}, D., \& {Plane}, J.~M.~C. 2016, \grl, 43, 11

\bibitem[{{Cintala}(1992)}]{Cintala_1992}
{Cintala}, M.~J. 1992, \jgr, 97, 947

\bibitem[{{Colaprete} {et~al.}(2010){Colaprete}, {Schultz}, {Heldmann},
  {Wooden}, {Shirley}, {Ennico}, {Hermalyn}, {Marshall}, {Ricco}, {Elphic},
  {Goldstein}, {Summy}, {Bart}, {Asphaug}, {Korycansky}, {Landis}, \&
  {Sollitt}}]{Colaprete_etal_2010}
{Colaprete}, A., {Schultz}, P., {Heldmann}, J., {et~al.} 2010, Science, 330,
  463

\bibitem[{Deutsch {et~al.}(2019)Deutsch, Head, \& Neumann}]{Deutsch_etal_2019}
Deutsch, A.~N., Head, J.~W., \& Neumann, G.~A. 2019, Icarus, 113455

\bibitem[{{Fisher} {et~al.}(2017){Fisher}, {Lucey}, {Lemelin}, {Greenhagen},
  {Siegler}, {Mazarico}, {Aharonson}, {Williams}, {Hayne}, {Neumann}, {Paige},
  {Smith}, \& {Zuber}}]{FISHER_ETAL_2017}
{Fisher}, E.~A., {Lucey}, P.~G., {Lemelin}, M., {et~al.} 2017, \icarus, 292, 74

\bibitem[{{Gault}(1973)}]{Gault_1973}
{Gault}, D.~E. 1973, Moon, 6, 32

\bibitem[{{Haruyama} {et~al.}(2013){Haruyama}, {Yamamoto}, {Yokota}, {Ohtake},
  \& {Matsunaga}}]{HARUYAMA_ETAL_2013}
{Haruyama}, J., {Yamamoto}, S., {Yokota}, Y., {Ohtake}, M., \& {Matsunaga}, T.
  2013, \grl, 40, 3814

\bibitem[{{Hayne} {et~al.}(2015){Hayne}, {Hendrix}, {Sefton-Nash}, {Siegler},
  {Lucey}, {Retherford}, {Williams}, {Greenhagen}, \&
  {Paige}}]{HAYNE_ETAL_2015}
{Hayne}, P.~O., {Hendrix}, A., {Sefton-Nash}, E., {et~al.} 2015, \icarus, 255,
  58

\bibitem[{{Hor{\'a}nyi} {et~al.}(2015){Hor{\'a}nyi}, {Szalay}, {Kempf},
  {Schmidt}, {Gr{\"u}n}, {Srama}, \& {Sternovsky}}]{Horanyi_etal_2015}
{Hor{\'a}nyi}, M., {Szalay}, J.~R., {Kempf}, S., {et~al.} 2015, \nat, 522, 324

\bibitem[{{Hurley} {et~al.}(2012){Hurley}, {Lawrence}, {Bussey}, {Vondrak},
  {Elphic}, \& {Gladstone}}]{Hurley_etal_2012}
{Hurley}, D.~M., {Lawrence}, D.~J., {Bussey}, D. B.~J., {et~al.} 2012, \grl,
  39, L09203

\bibitem[{{Janches} {et~al.}(2015){Janches}, {Close}, {Hormaechea},
  {Swarnalingam}, {Murphy}, {O'Connor}, {Vandepeer}, {Fuller}, {Fritts}, \&
  {Brunini}}]{Janches_etal_2015}
{Janches}, D., {Close}, S., {Hormaechea}, J.~L., {et~al.} 2015, \apj, 809, 36

\bibitem[{{Koschny} \& {Gr{\"u}n}(2001)}]{Koschny_Grun_2001}
{Koschny}, D., \& {Gr{\"u}n}, E. 2001, \icarus, 154, 402

\bibitem[{{Lemelin} {et~al.}(2016){Lemelin}, {Lucey}, {Neumann}, {Mazarico},
  {Barker}, {Kakazu}, {Trang}, {Smith}, \& {Zuber}}]{Lemelin_etal_2016}
{Lemelin}, M., {Lucey}, P.~G., {Neumann}, G.~A., {et~al.} 2016, \icarus, 273,
  315

\bibitem[{{Li} {et~al.}(2018){Li}, {Lucey}, {Milliken}, {Hayne}, {Fisher},
  {Williams}, {Hurley}, \& {Elphic}}]{Li_etal_2018}
{Li}, S., {Lucey}, P.~G., {Milliken}, R.~E., {et~al.} 2018, Proceedings of the
  National Academy of Science, 115, 8907

\bibitem[{{Li} \& {Milliken}(2017)}]{Li_Milliken_2017}
{Li}, S., \& {Milliken}, R.~E. 2017, Science Advances, 3, e1701471

\bibitem[{{Lucey} {et~al.}(2014){Lucey}, {Neumann}, {Riner}, {Mazarico},
  {Smith}, {Zuber}, {Paige}, {Bussey}, {Cahill}, {McGovern}, {Isaacson},
  {Corley}, {Torrence}, {Melosh}, {Head}, \& {Song}}]{Lucey_etal_2014}
{Lucey}, P.~G., {Neumann}, G.~A., {Riner}, M.~A., {et~al.} 2014, Journal of
  Geophysical Research (Planets), 119, 1665

\bibitem[{{Mazarico} {et~al.}(2018){Mazarico}, {Barker}, \&
  {Nicholas}}]{Mazarico_etal_2018}
{Mazarico}, E., {Barker}, M.~K., \& {Nicholas}, J.~B. 2018, Advances in Space
  Research, 62, 3214

\bibitem[{{Morgan} \& {Shemansky}(1991)}]{Morgan_Shemansky_1991}
{Morgan}, T.~H., \& {Shemansky}, D.~E. 1991, \jgr, 96, 1351

\bibitem[{{Nesvorn{\'y}} {et~al.}(2011){Nesvorn{\'y}}, {Vokrouhlick{\'y}},
  {Pokorn{\'y}}, \& {Janches}}]{Nesvorny_etal_2011OCC}
{Nesvorn{\'y}}, D., {Vokrouhlick{\'y}}, D., {Pokorn{\'y}}, P., \& {Janches}, D.
  2011, \apj, 743, 37

\bibitem[{{Paige} {et~al.}(2010){Paige}, {Foote}, {Greenhagen}, {Schofield},
  {Calcutt}, {Vasavada}, {Preston}, {Taylor}, {Allen}, \&
  {Snook}}]{Paige_etal_2010}
{Paige}, D.~A., {Foote}, M.~C., {Greenhagen}, B.~T., {et~al.} 2010, \ssr, 150,
  125

\bibitem[{{Pokorn{\'y}} {et~al.}(2019){Pokorn{\'y}}, {Janches}, {Sarantos},
  {Szalay}, {Hor{\'a}nyi}, {Nesvorn{\'y}}, \& {Kuchner}}]{Pokorny_etal_2019}
{Pokorn{\'y}}, P., {Janches}, D., {Sarantos}, M., {et~al.} 2019, Journal of
  Geophysical Research (Planets), 124, 752

\bibitem[{{Pokorn{\'y}} {et~al.}(2014){Pokorn{\'y}}, {Vokrouhlick{\'y}},
  {Nesvorn{\'y}}, {Campbell-Brown}, \& {Brown}}]{Pokorny_etal_2014}
{Pokorn{\'y}}, P., {Vokrouhlick{\'y}}, D., {Nesvorn{\'y}}, D.,
  {Campbell-Brown}, M., \& {Brown}, P. 2014, \apj, 789, 25

\bibitem[{{Rubanenko} {et~al.}(2019){Rubanenko}, {Venkatraman}, \&
  {Paige}}]{Rubanenko_etal_2019}
{Rubanenko}, L., {Venkatraman}, J., \& {Paige}, D.~A. 2019, Nature Geoscience,
  12, 597

\bibitem[{{Smith} {et~al.}(2010){Smith}, {Zuber}, {Neumann}, {Lemoine},
  {Mazarico}, {Torrence}, {McGarry}, {Rowlands}, {Head}, \&
  {Duxbury}}]{Smith_etal_2010}
{Smith}, D.~E., {Zuber}, M.~T., {Neumann}, G.~A., {et~al.} 2010, \grl, 37,
  L18204

\bibitem[{{Smith} {et~al.}(2017){Smith}, {Zuber}, {Neumann}, {Mazarico},
  {Lemoine}, {Head}, {Lucey}, {Aharonson}, {Robinson}, {Sun}, {Torrence},
  {Barker}, {Oberst}, {Duxbury}, {Mao}, {Barnouin}, {Jha}, {Rowlands},
  {Goossens}, {Baker}, {Bauer}, {Gl{\"a}ser}, {Lemelin}, {Rosenburg}, {Sori},
  {Whitten}, \& {Mcclanahan}}]{Smith_etal_2017}
---. 2017, \icarus, 283, 70

\bibitem[{{Suggs} {et~al.}(2014){Suggs}, {Moser}, {Cooke}, \&
  {Suggs}}]{Suggs_etal_2014}
{Suggs}, R.~M., {Moser}, D.~E., {Cooke}, W.~J., \& {Suggs}, R.~J. 2014,
  \icarus, 238, 23

\bibitem[{{Tye} {et~al.}(2015){Tye}, {Fassett}, {Head}, {Mazarico},
  {Basilevsky}, {Neumann}, {Smith}, \& {Zuber}}]{TYE_ETAL_2015}
{Tye}, A.~R., {Fassett}, C.~I., {Head}, J.~W., {et~al.} 2015, \icarus, 255, 70

\bibitem[{{Zuber} {et~al.}(2012){Zuber}, {Head}, {Smith}, {Neumann},
  {Mazarico}, {Torrence}, {Aharonson}, {Tye}, {Fassett}, {Rosenburg}, \&
  {Melosh}}]{Zuber_etal_2012}
{Zuber}, M.~T., {Head}, J.~W., {Smith}, D.~E., {et~al.} 2012, \nat, 486, 378

\end{thebibliography}

\end{document}